Fermilab

# Study of secondary neutron interactions with $^{232}$Th, $^{129}$I, and $^{127}$I nuclei with the uranium assembly "QUINTA " at 2, 4, and 8 GeV deuteron beams of the JINR Nuclotron accelerator[1][2]

J. Adam[1,2], V.V. Chilap[3], V.I. Furman[1], M.G. Kadykov[1], J. Khushvaktov[1], V.S. Pronskikh[1,4], A.A. Solnyshkin[1], V.I. Stegailov[1], M. Suchopar[2], V.M. Tsoupko-Sitnikov[1], S.I. Tyutyunnikov[1], J. Vrzalova[1], V. Wagner[2], L. Zavorka[1]

[1]*Joint Institute for Nuclear Research, Dubna, Russia.*
[2]*Nuclear Physics Institute ASCR PRI, Czech Republic.*
[3]*Center of Physical and Technical Projects "Atomenergomash", Moscow, Russia.*
[4]*Fermi National Accelerator Laboratory, Batavia IL, USA*

Abstract

The natural uranium assembly, "QUINTA", was irradiated with 2, 4, and 8 GeV deuterons. The 232 Th, 127 I, and 129 I samples have been exposed to secondary neutrons produced in the assembly at a 20-cm radial distance from the deuteron beam axis. The spectra of gamma rays emitted by the activated 232 Th, 127 I, and 129 I samples have been analyzed and several tens of product nuclei have been identified. For each of those products, neutron-induced reaction rates have been determined. The transmutation power for the 129 I samples is estimated. Experimental results were compared to those calculated with well-known stochastic and deterministic codes.

[1] Work supported by Fermi Research Alliance, LLC under contract No. DE-AC02-07CH11359 with the U.S. Department of Energy.
[2] Accepted to Applied Radiation and Isotopes, 2015.

# Study of secondary neutron interactions with $^{232}Th$, $^{129}I$, and $^{127}I$ nuclei with the uranium assembly " QUINTA " at 2, 4, and 8 GeV deuteron beams of the JINR Nuclotron accelerator

J. Adam[1,2], V.V. Chilap[3], V.I. Furman[1], M.G. Kadykov[1], J. Khushvaktov[1], V.S. Pronskikh[1,4], A.A. Solnyshkin[1], V.I. Stegailov[1], M. Suchopar[2], V.M. Tsoupko-Sitnikov[1], S.I. Tyutyunnikov[1], J. Vrzalova[1], V. Wagner[2], L. Zavorka[1],

[1]*Joint Institute for Nuclear Research, Dubna, Russia.*

[2]*Nuclear Physics Institute ASCR PRI, Czech Republic.*

[3]*Center of Physical and Technical Projects "Atomenergomash", Moscow, Russia.*

[4]*Fermi National Accelerator Laboratory, Batavia IL, USA*

## INTRODUCTION

Interest in the international scientific community for research of this kind, is primarily concerned with the problem of transmutation of long-lived radioactive waste [1,2] and the creation of subcritical nuclear power plants with uranium-thorium cycle, controlled high-energy particle accelerators (Accelerator Driven Subcritical Systems) [3,4]. Such research is actively conducted throughout the world has been for the last two decades: PNF (Poahng) [5], n-ToF (CERN) [6], MYRRHA (Belgium) [7] and «Energy + Transmutation» setup at JINR (Dubna) [8, 9, 10, 11]. Currently working in this direction and develops a number of programs: SINQ (PSI) [12], KEK (Japan) [13], MYRRHA (Belgium), n-ToF (CERN) and a cluster of other research programs at LANL (USA) [14] – for obtaining data and developing new materials to create prototypes industrial ADS-systems.

During the past several years, such studies have been conducted and are ongoing with beams of particles Nuclotron VBLHEP JINR (Dubna) under the program Energy plus Transmutation of Radioactive Waste. This program was carried out a large number of experiments on subcritical uranium-lead target "QUINTA" [15, 16], as well as lead-graphite target "GAMMA-3" [17]. Several experiments were carried out using a solid lead target "GENERATOR" [18, 19, 20] with proton beam Phasotron DLNP JINR.

In this paper we present experimental data in comparison with the calculations obtained in the last two years in studying the interaction of secondary neutrons with nucleus $^{232}$Th, $^{129}$I, $^{127}$I on the "QUINTA" VBLHEP JINR on deuteron beams with energies 2, 4, 8 GeV.

## STRUCTURE SETUP "QUINTA"

Uranium assembly "Quinta" is presented in Fig.1. It consists of five sections, formed in the shape of a hexahedron (aluminum containers with an inscribed diameter of 284 mm). Containers filled cylindrical rods of natural uranium metal, having a sealed aluminum shell (external dimensions: diameter 3.6 cm, length 10.4 cm, weight 1.72 kg of uranium). The ends are made of aluminum sections, 6 mm thick. The first section, the first located along the beam contains 54 uranium rod and has a through central opening 80 mm in diameter for the input beam into the target, made in order to reduce its albedo and reduce the leakage of neutrons from the target. Four subsequent sections are structurally identical and contain 61 uranium rods. Mass of uranium in one section is 61x1.72=104.92 kg, and the total mass of uranium entire target 298x1.72=512.56 kg. The fill factor of uranium 2, 3, 4 or 5 sections about 0.8, and the whole assembly of uranium ~ 0.6.

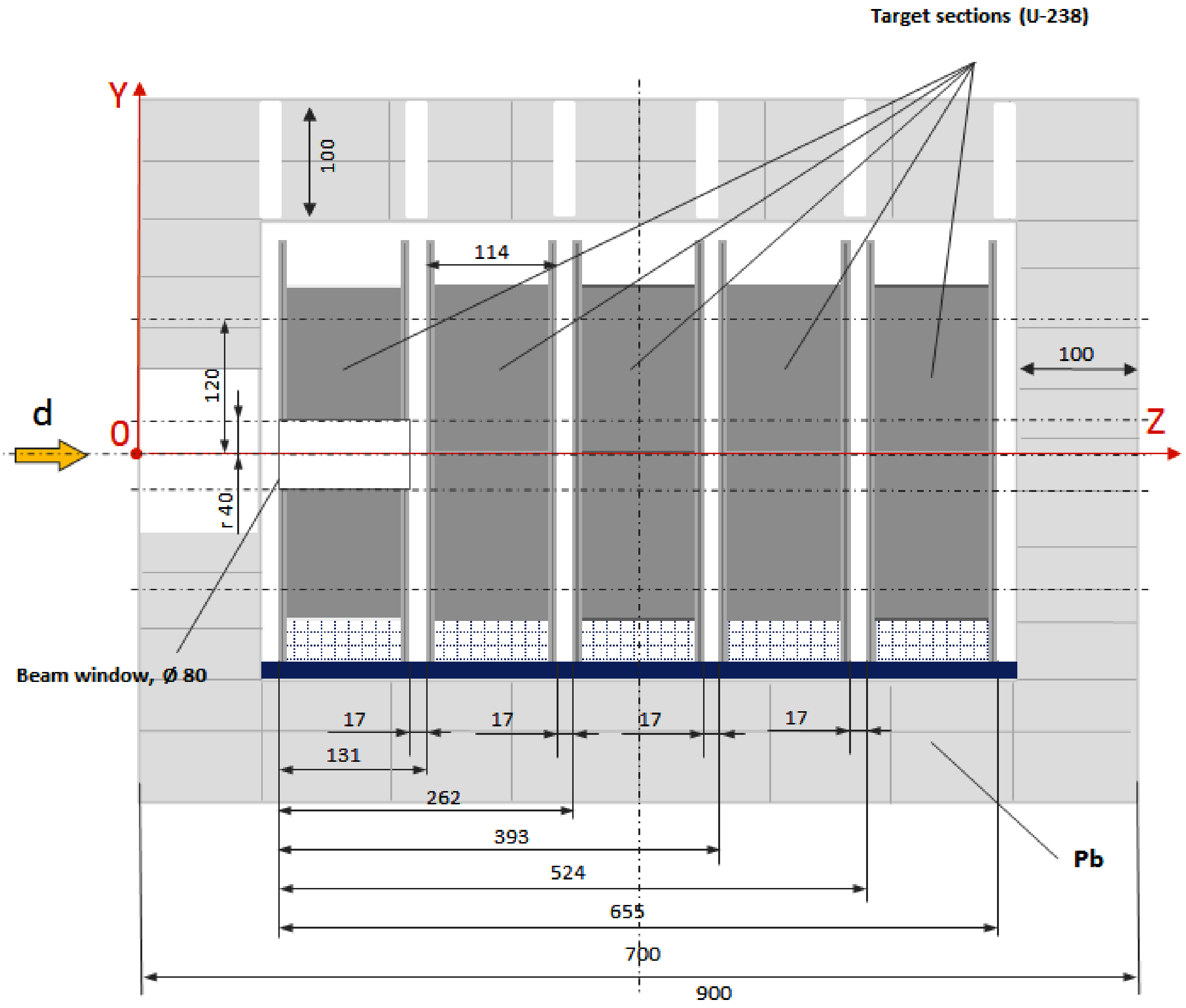

Fig. 1. The general scheme of the setup "QUINTA"

Uranium target surrounded with lead shield thickness of 10 cm and a weight of 2545 kg with a window to enter the beam dimensions of 15 x 15 $cm^2$ (see Fig. 2). In the side wall of lead shield on opposite to the third section there is window hole in the size 15 x 5 $cm^2$ to accommodate transmutation samples. The upper part of the lead assembly is provided with a special hole for mounting and dismantling of detector probes and installation of the samples inside the uranium assemblies between sections.

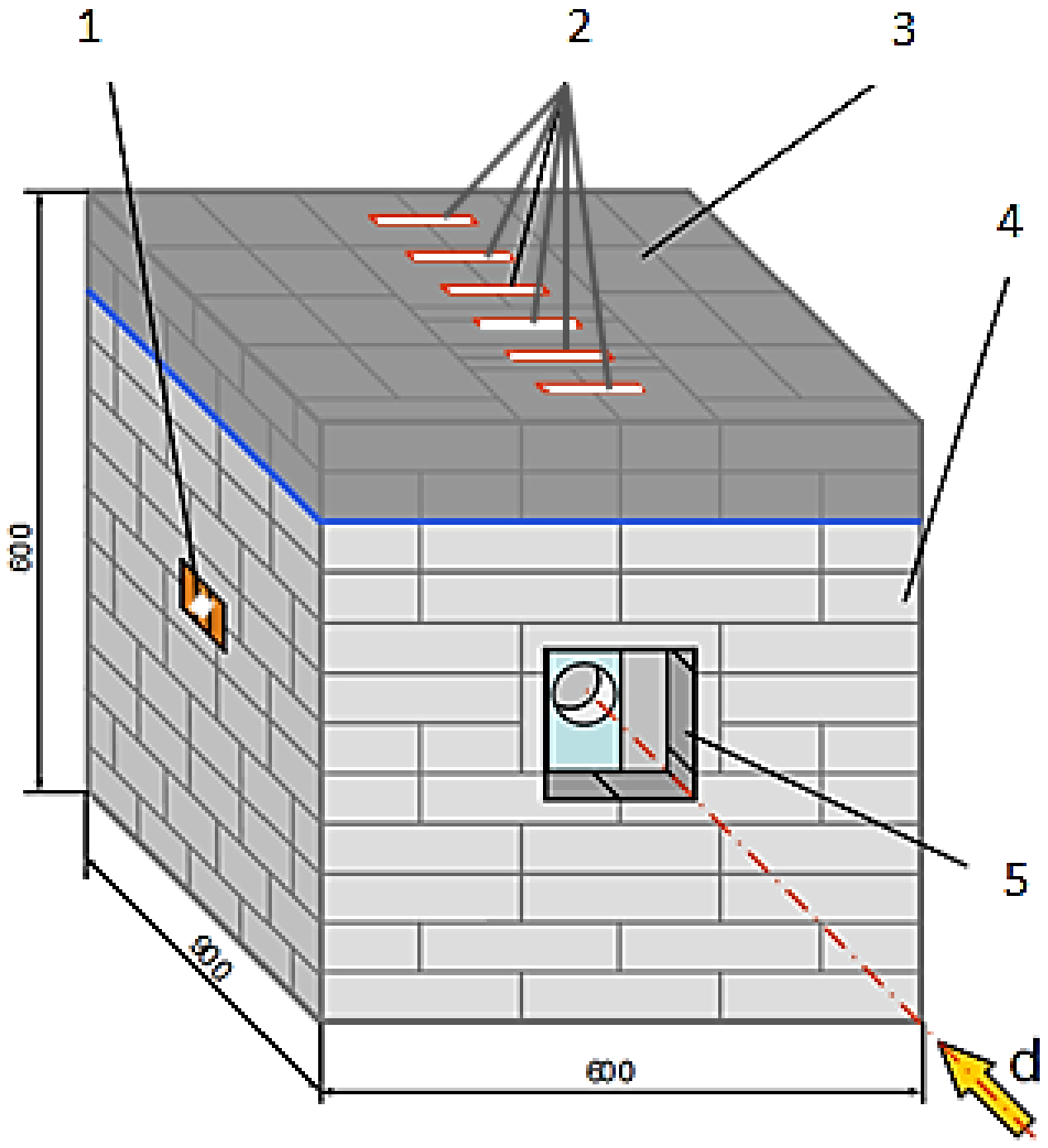

Fig. 2. The general view of setup "QUINTA", where 1 – window for placement of transmutation samples, 2 - mine for mounting and dismantling of detector probes and installation of the samples inside the uranium assemblies between sections, 3 - cover lead assembly, 4 - lead assembly, 5 - input box beam 15x15 $cm^2$.

## EXPERIMENT

In our experiments, transmutation samples ($^{127}I$, $^{129}I$, $^{nat}Th$, $^{233}U$, $^{235}U$, $^{nat}U$, $^{237}Np$, $^{238}Pu$, $^{239}Pu$, and $^{241}Am$) were placed inside the window 1 (see Fig. 2). Irradiation was carried out on three deuteron energies (2, 4 and 8 GeV). Before entering the deuteron beam at the target were installed aluminum and copper foil. It is possible to determine the integral flux of deuterons and beam shape. Used the widely known method of activation of aluminum foil in the reaction $^{27}Al(d,x)^{24}Na$ and copper foil in the reaction $^{nat}Cu(d,x)^{24}Na$. The following Table 1 shows the irradiation conditions and the characteristics of the samples that were used in our studies ($^{nat}Th$, $^{129}I$ and $^{127}I$).

Table.1. Data on the irradiation conditions and characteristics of the samples.

| Deuterons energy, GeV | 2 | | 4 | | 8 | |
|---|---|---|---|---|---|---|
| Irradiation time, min. | 376 | | 561 | | 970 | |
| Integral number of deuterons | 3.02(10)E+13 | | 2.73(10)E+13 | | 9.10(40)E+12 | |
| Coordinates of the center of beam, cm * | Xc | Yc | Xc | Yc | Xc | Yc |
| | 1.5(2) | 0.1(1) | 1.8(1) | -0.3(1) | 0.9(1) | 0.1(1) |
| FWHM (full width at half maximum), cm * | $FWHM_X$ | $FWHM_Y$ | $FWHM_X$ | $FWHM_Y$ | $FWHM_X$ | $FWHM_Y$ |
| | 2.0(1) | 1.7(2) | 1.5(2) | 1.1(1) | 1.0(1) | 1.3(1) |
| Samples | Th-nat | | Th-nat | | Th-nat | |
| Mass, g. | 0.975 | | 1.000 | | 0.249 | |
| Diameter of samples, cm | 1.3 | | 1.3 | | 1.3 | |
| Samples | I-129 | I-127 | I-129 | I-127 | I-129 | I-127 |
| Mass (I-129), g. | 0.591 | - | 0.339 | - | 0.218 | - |
| Mass (I-127), g. | 0.129 | 1.550 | 0.074 | 1.270 | 0.048 | 1.980 |
| Mass (Na-23), g. | 0.118 | 0.290 | 0.067 | 0.230 | 0.043 | 0.360 |
| Mass (Al-27), g. ** | 17.6 | - | 17.6 | - | 17.6 | - |
| Diameter of samples, cm | 2.1 | 2.0 | 2.1 | 2.0 | 2.1 | 2.0 |

* Deuteron beam parameters (plata0) from group I.V. Zhuk (Minsk). Determined using solid-state nuclear track detectors with radiators from a natural lead, called sensors [private message].

** Container for radioactive $^{129}$I.

After each session of irradiation the studying samples were transported to the DLNP by complex YASNAPP-2 where measured γ-spectra with the three spectrometers based HPGe-detector ORTEC (single detector efficiency - 33% energy resolution - 1.8 keV at line 1.33 MeV $^{60}$Co) and CANBERRA (two detector efficiency - 18% and 30%, the energy resolutions - 1.9 keV and 1.8 keV 1.33 MeV at line $^{60}$Co). For each sample at various time intervals was measured from 13 to 16 γ-spectra and decay time to the first spectrum from 79 to 157 min. Calibration of the detectors by energy and efficiency was performed using a standard set of sources ($^{54}$Mn, $^{57}$Co, $^{60}$Co, $^{88}$Y, $^{113}$Sn, $^{133}$Ba, $^{137}$Cs, $^{139}$Ce, $^{152}$Eu, $^{228}$Th, $^{241}$Am).

## PROCESSING OF γ- SPECTRA

Processing of the measured γ-spectra were using DEIMOS32 [21]. The program allows to define the area under the γ-peaks and their position (channel number). Next, using a special software package [22] was calibrated for energy, corrects for detector efficiency and identified separate γ-lines corresponding nuclei products that were emitted in the sample as a result of interaction with secondary neutrons. In determining the intensities of γ-transitions also introduced corrections to the nuclear decay products during irradiation, corrections for self-absorption for registered γ-rays in the sample, to the geometric dimensions of the sample, corrections to breaks during irradiation and the change in intensity of the deuteron beam (on-line measurements of fast ionization chambers). All these procedures are described in detail in [9, 22, 23].

## METHOD OF ANALYSIS AND CALCULATIONS

In determining the reaction rates (experimental data) used the following relation (1) [9]:

$$R(A_r, Z_r) = \frac{Q(A_r, Z_r)}{N_t N_d} \quad , \tag{1}$$

where, $Q(A_r, Z_r)$ – rate of production of radioactive nucleus (r), $N_t$ – number of atoms in the sample, $N_d$ - number of incident deuterons on the target.

Values of reaction rates (in calculation) calculated by the formula:

$$R(A_r, Z_r) = \int_{E_{thr}(A_r, Z_r)}^{\infty} \sigma(A_r, Z_r, E_n)\varphi(E_n)dE_n \tag{2}$$

where, $\sigma(A_r, Z_r, E_n)$ – reaction cross section, $\varphi(E_n)$ – neutron fluence.

Calculations of reaction rates (Calc.1) were performed with program MARS15 [24] and the reaction products with neutrons were modeled using LAQGSM03.03 [25].

Calculation of neutron fluence (Calc.2) was carried with program MCNPX2.7 [26] using models INCL4 (intranuclear physics model) [27] and ABLA (fission-evaporation model) [28]. For the reaction (n,γ) in the $^{232}$Th, the reaction cross sections were calculated by the program TALYS1.4 [29], and (n,fission) were taken from the nuclear data library TENDL-2009 [30], as in the TENDL-2009, thorium fission reaction cross sections with neutrons calculated up to neutron energy 200 MeV and are in good agreement with the data from the library JEFF3.1.2 [31], as well as with the experimental data [32] (see Figure.3). Reaction cross sections for (n,γ), (n,4n) and (n,6n) in $^{129}$I to 40 MeV from TENDL-2011, from 40 to 200 MeV, the calculation was performed with program TALYS1.4. In $^{127}$I reaction cross sections for (n,γ), (n,2n) and (n,4n) chosen the calculations with program TALYS1.4 lack of data in libraries up to 200 MeV.

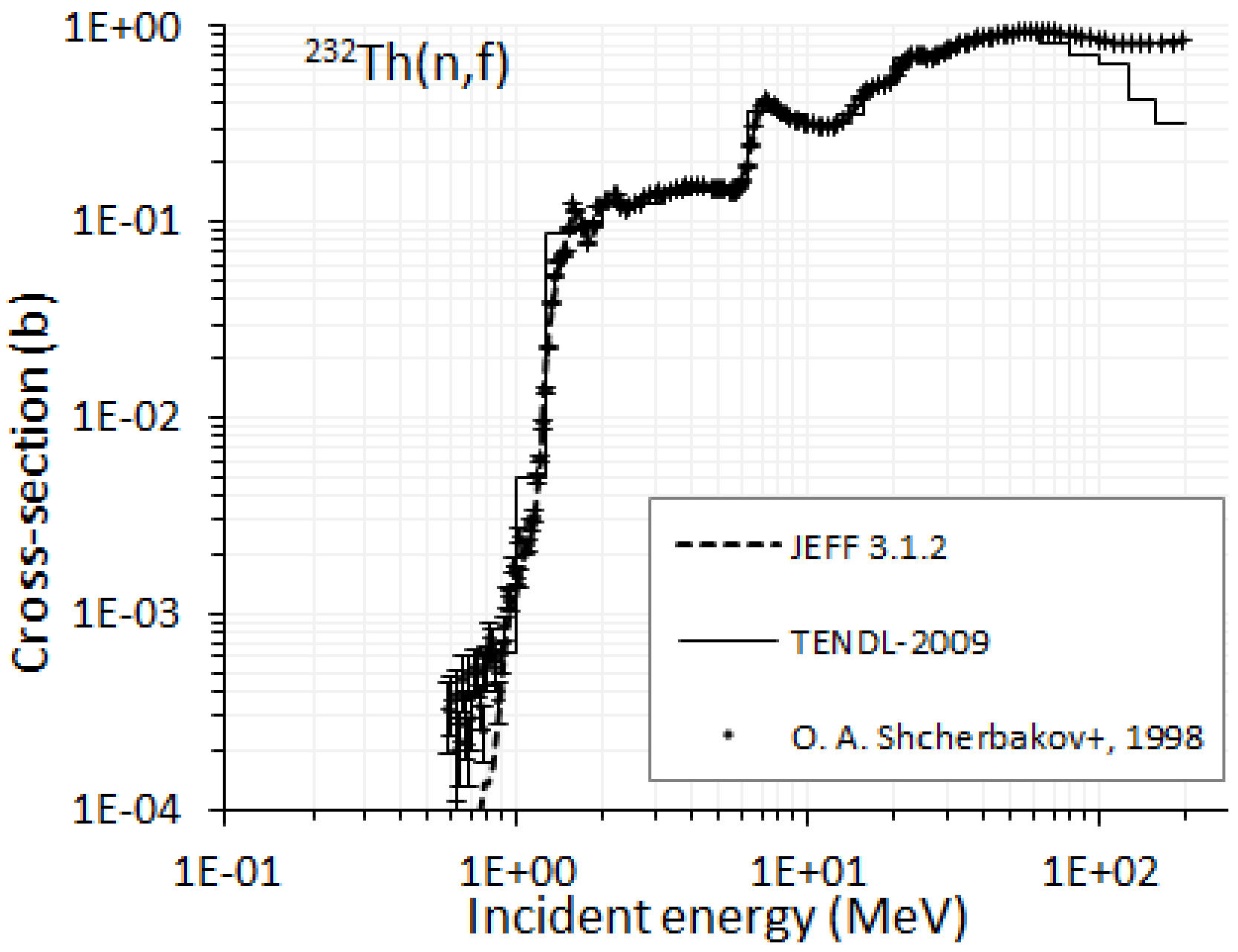

Fig.3. Comparison of the calculated fission cross sections of $^{232}$Th with experiment [32], depending on the neutron energy.

## RESULTS OF $^{232}$Th

In results processing of obtained data in the experiment for thorium has been identified a large number of product nuclei (at 2 GeV – 19, at 4 GeV - 30 and at 8 GeV - 27) for three values of the deuteron energy. For each of these are obtained reaction rates. Table 2 summarizes the results of $^{232}$Th for all energies for all registered nuclei. $I_g$ – the yield of gamma rays (%), $T_{1/2}$(Exper) – the observed half-lives of the radionuclides, R – reaction rate, <R> - the average value of the reaction rate (atoms$^{-1}$ * deuteron$^{-1}$). The results show that with increasing deuteron energy increases and the value of the reaction rates for almost all product nuclei. Obviously, this growth is due to the increased flow and energy of secondary neutrons with increasing energy deuterons. Fig.4 shows the ratio of reaction rate R(4 GeV) / R(2 GeV) and R(8 GeV) / R(2 GeV) for the identified product nuclei generated in reactions with secondary neutrons at all three deuteron energies (2, 4, 8 GeV).

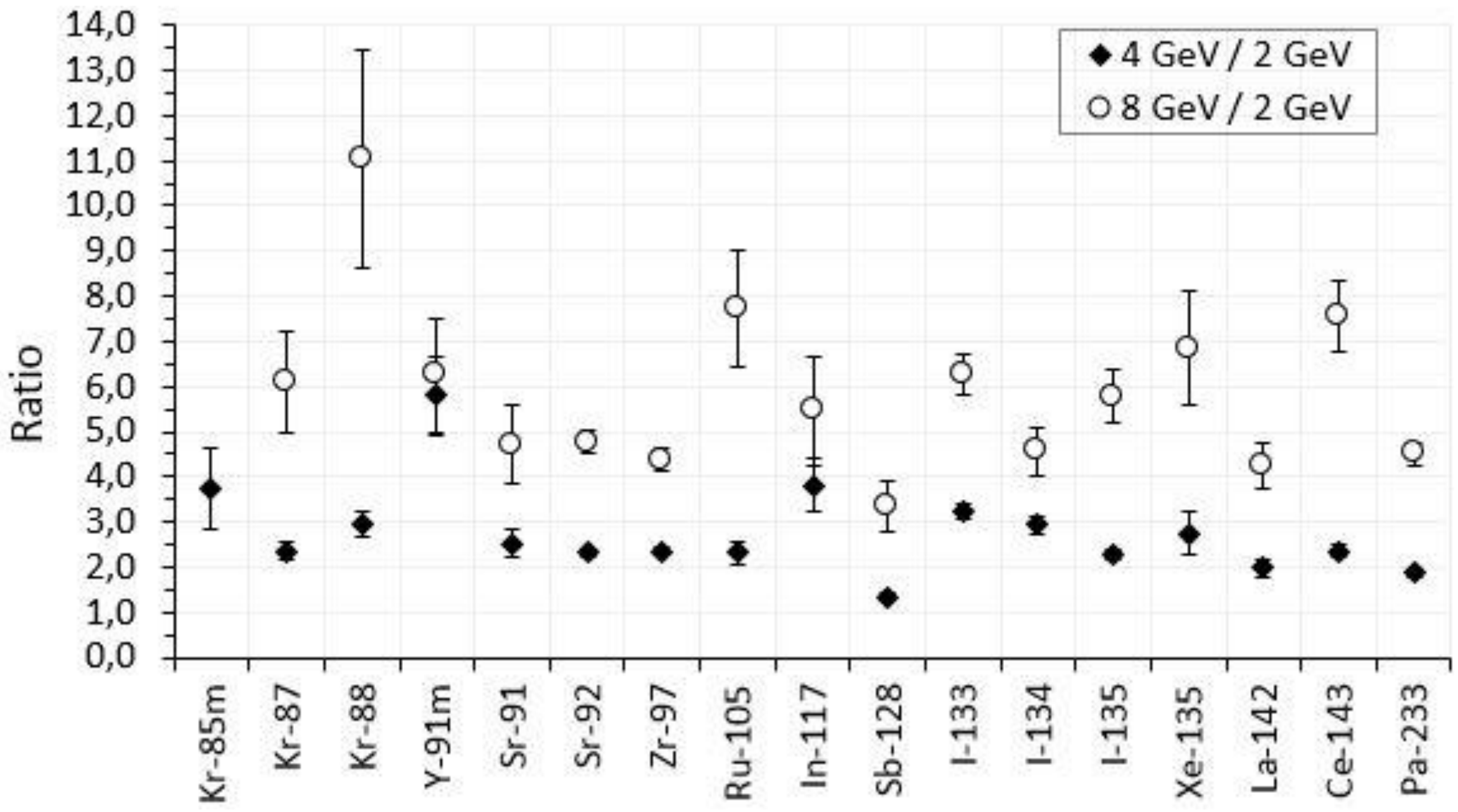

Fig.4. Experimental values of the ratio of reaction rate R(4 GeV) / R(2 GeV) and R(8 GeV) / R(2 GeV) for $^{232}$Th with secondary neutrons for product nuclei at energies of deuterons 2, 4, 8 GeV.

Presented in Fig.4 nuclei products: $^{85m}$Kr, $^{87}$Kr, $^{88}$Kr, $^{91m}$Y, $^{91}$Sr, $^{92}$Y, $^{92}$Sr, $^{93}$Y, $^{97}$Nb, $^{97}$Zr, $^{133}$I, $^{134}$I, $^{135}$I, $^{135}$Xe, $^{138}$Cs, $^{142}$La and $^{143}$Ce – produced by fission the $^{232}$Th; radionuclides: $^{66}$Ga, $^{88}$Y, $^{92m}$Nb, $^{105}$Ru, $^{115}$Cd, $^{115m}$In, $^{117}$In $^{126}$Sb, $^{128}$Sb, $^{129}$Sb, $^{132}$Te и $^{132}$I – products of the (n,spallation) reactions (the ratio of the reaction rate to yield R / Y for these radionuclides 10-20 times differed from fission products). $^{224}$Ac is product of reaction $^{228}$Th(n,2nt)$^{224}$Ac ($E_{thr}$ = 16.56 MeV), $^{233}$Pa produced in the reaction $^{232}$Th(n,γ)$^{233}$Th ($\beta^-$ decay, $T_{1/2}$=22.3 min.) → $^{233}$Pa ($\beta^-$ decay, $T_{1/2}$=26.967 day) → $^{233}$U.

Fig.5 shows the experimental values of the reaction rate (n,γ) and fission (n,f) depending on the deuteron energy in comparison with the calculations Calc.2 (MCNPX). In both cases, there is a proportional increase in the values of reaction rate with increasing energy deuterons.

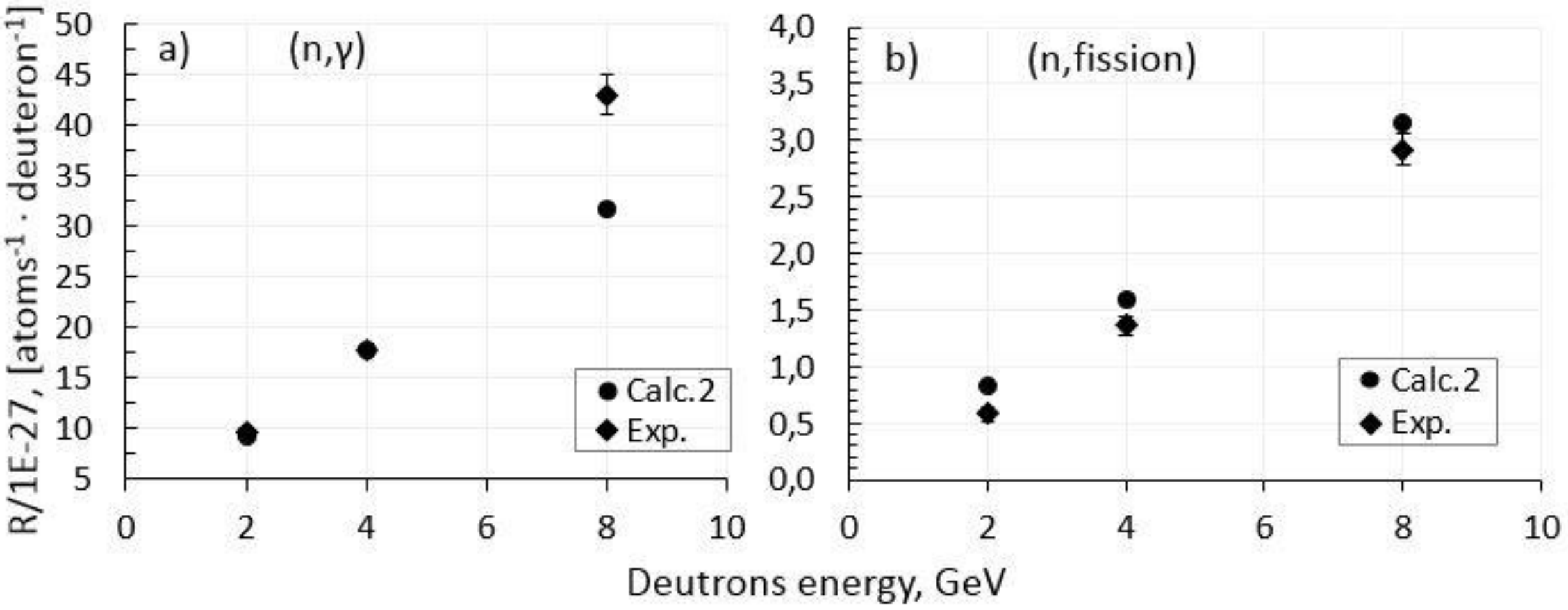

Fig.5. a) Comparison of experimental and calculated values of the reaction rate (n,γ) $^{232}$Th in the interaction with the secondary neutrons, depending from deuteron energy, b) Comparison of experimental and calculated values of the fission reaction rate (n,f) in the interaction of $^{232}$Th with secondary neutrons depending from deuteron energy.

Calculations of fission reaction rates R(n,f) from the experimental data were carried out as follows. Cumulative yield Y for fission products of $^{232}$Th were taken from the library JEFF3.1 for neutrons with an energy 0.4 MeV. The average value of the ratio R/Y for fission products such as nuclei $^{85m}$Kr, $^{87}$Kr, $^{88}$Kr, $^{91m}$Y, $^{91}$Sr, $^{92}$Y, $^{92}$Sr, $^{93}$Y, $^{97}$Nb, $^{97}$Zr, $^{133}$I, $^{134}$I, $^{135}$I, $^{135}$Xe, $^{138}$Cs, $^{142}$La and $^{143}$Ce in experiment for deuteron energies at 2 GeV is 0.58(6)E-27, at 4 GeV is 1.36(9)E-27 and at 8 GeV is 2.92(14)E-27. These numbers is the fission reaction rates R(n,f) for $^{232}$Th.

On the experimental results values of ratio (n,γ)/(n,f) for different energies of deuterons: 2 GeV – 16.4(24), at 4 GeV – 13.0(20), at 8 GeV – 14.8(22). Calculated values (Cal.2.) for these ratios (n,γ)/(n,f) are at 2 GeV - 12.3, at 4 GeV - 12.2 and at 8 GeV - 11.2.

Table.2. Values of the reaction rates $^{232}Th$ with secondary neutrons for product nuclei at energies of deuterons 2, 4, 8 GeV.
(*) denotes mixing due to other nuclide.

| **Isotope** Energy [keV] | $I_g$ [%] | **2 GeV** | | **4 GeV** | | **8 GeV** | |
|---|---|---|---|---|---|---|---|
| | | **$T_{1/2}$(Library)** $T_{1/2}$(Exper.) | **<R>** R | **$T_{1/2}$(Library)** $T_{1/2}$(Exper.) | **<R>** R | **$T_{1/2}$(Library)** $T_{1/2}$(Exper.) | **<R>** R |
| **Ga-66** | | | | **9.49(7) h** | | **9.49(7) h** | |
| 1039.231 | 37 | | | 12.6(20) h | **2.68(62)E-29** | 6(5) h | **1.05(23)E-28** |
| **Kr-85m** | | **4.48(1) h** | | **4.48(1) h** | | | |
| 151.159 | 75 | 7(3) d | **1.47(48)E-29** | | | | |
| 304.870 | 14 | | | 3.2(11) h | **5.52(85)E-29** | | |
| **Kr-87** | | **76.3(6) m** | | **76.3(6) m** | | **76.3(6) m** | |
| 402.586 | 49.6 | 1.08(12) h | **4.32(47)E-29** | 1.29(19) h | **1.02(7)E-28** | 1.7(7) h | **2.64(69)E-28** |
| **Kr-88** | | **2.84(3) h** | **2.90(23)E-29** | **2.84(3) h** | **8.61(94)E-29** | **2.84(3) h** | |
| 196.301 | 26 | 2.3(4) h | 2.85(25)E-29 | | | | |
| 1529.770 | 10.9 | | | 3.9(1) h | 1.08(11)E-28 | 12(5) h | **3.2(11)E-28** |
| 2195.842 | 13.2 | | 3.36(71)E-29 | | 7.4(10)E-29 | | |
| 2392.110 | 34.6 | 2.5(22) h | 3.22(95)E-29 | 2.11(22) h | 8.17(93)E-29 | | |
| **Y-88** | | | | **106.65(4) d** | | **106.65(4) d** | |
| 898.042 | 93.7 | | | | | | |
| 1836.063 | 99.2 | | | | **2.99(36)E-28** | | **1.11(18)E-27** |
| **Y-91mD** | | **49.71(4) m** | | **49.71(4) m** | | **49.71(4) m** | |
| 555.570 | 95 | 10.3(12) h | **2.78(68)E-29** | | **1.61(9)E-28** | | **1.74(27)E-28** |
| **Sr-91** | | **9.63(5) h** | **4.00(90)E-29** | **9.63(5) h** | **1.01(3)E-28** | **9.63(5) h** | **1.89(29)E-28** |
| 652.900 | 8 | | | 14(10) h | 1.15(11)E-28 | | |
| 749.800 | 23.6 | 10.8(10) h | 4.85(25)E-29 | 9.1(7) h | 1.01(4)E-28 | 1.15(25) d | 2.09(31)E-28 |
| 1024.300 | 33 | 8.8(15) h | 2.82(29)E-29 | 9.9(4) h 7 | 1.01(3)E-28 | 9.1(16) h | 1.83(26)E-28 |
| **Sr-92** | | **2.71(1) h** | | **2.71(1) h** | | **2.71(1) h** | |
| 1383.930 | 90 | 2.62(10) h | **3.81(16)E-29** | 2.68(11) h | **8.81(32)E-29** | 2.74(27) h | **1.82(13)E-28** |
| **Y-92** | | | | **3.54(1) h** | **1.45(13)E-28** | **3.54(1) h** | **4.7(10)E-28** |
| 934.460 | 13.9 | | | | 1.63(36)E-28 | | 4.4(14)E-28 |
| 1405.280 | 4.8 | | | 5.1(12) h | 1.42(14)E-28 | | 4.9(16)E-28 |
| **Nb-92m** | | | | **10.15(2) d** | | **10.15(2) d** | |
| 934.460 | 99 | | | | **2.43(19)E-28** | | **3.16(69)E-28** |
| **Y-93** | | **10.18(8) h** | | | | | |
| 266.900 | 7.3 | 9(3) h | **4.28(56)E-29** | | | | |
| **Zr-97** | | **16.91(5) h** | | **16.91(5) h** | | **16.91(5) h** | |
| 743.360 | 93 | 14.7(8) h | **2.88(14)E-29** | 15.7(6) h | **6.76(18)E-29** | 17.2(24) h | **1.26(9)E-28** |
| **Nb-97** | | | | **72.1(7) m** | | | |
| 658.080 | 98 | | | | **6.89(46)E-29** | | |
| **Ru-105** | | **4.44(2) h** | | **4.44(2) h** | **2.66(28)E-29** | **4.44(2) h** | |
| 469.370 | 17.5 | | | | 2.03(78)E-29 | | |
| 724.210 | 47 | 9(6) h | **1.15(11)E-29** | 5.7(26) h | 2.74(30)E-29 | | **8.9(22)E-29** |
| **Cd-115** | | | | **53.46(1) h** | | **53.46(1) h** | |
| 336.240 | 45.9 | | | | | | |
| 527.900 | 27.5 | | | 1.7(5) d | **3.88(62)E-29** | | **1.98(65)E-28** |
| **In-115m** | | **4.87(1) h** | | | | | |
| 336.240 | 45.8 | 4.4(20) h | **8.4(15)E-30** | | | | |
| **In-117D** | | **43.2(3) m** | **1.90(34)E-29** | **43.2(3) m** | | **43.2(3) m** | |
| 158.562 | 87 | | 1.69(37)E-29 | | | | |
| 553.000 | 100 | | 2.12(32)E-29 | | **7.26(95)E-29** | 26.5(1) m | **1.04(27)E-28** |
| **Sb-126** | | | | **12.46(3) d** | **4.81(51)E-29** | **12.46(3) d** | **3.74(52)E-28** |
| 414.810 | 83.3 | | | | | | |
| 666.331 | 100 | | | | 4.0(11)E-29 | | 4.38(77)E-28 |
| 695.030 | 100 | | | | 5.06(58)E-29 | | 3.03(88)E-28 |
| **Sn-127** | | | | | | **91.1(5) m** | |
| 1114.300 | 39 | | | | | 2.08(1) h | **6.9(20)E-29** |
| **Sb-128** | | **9.01(3) h** | | **9.01(3) h** | | **9.01(3) h** | **7.1(16)E-29** |
| 314.120 | 61 | | | | | | 4.6(23)E-29 |
| 526.570 | 45 | | | 4.5(22) h | **2.76(21)E-29** | | 7.9(22)E-29 |
| 743.220 | 100 | 14.7(8) h | **2.08(10)E-29** | | 5.56(59)E-29* | | 8.0(14)E-29 |
| **Sb-129** | | | | **4.40(1) h** | | | |
| 812.800 | 43 | | | 6(4) h | **7.9(16)E-30** | | |
| **Te-132** | | | | **3.20(1) d** | | **3.20(1) d** | |
| 228.160 | 88 | | | 15(10) d | **4.64(94)E-29** | | **1.46(38)E-28** |
| **I-132** | | | | **2.29(1) h** | **1.49(65)E-29** | | |
| 667.718 | 99 | | | | 1.35(77)E-29 | | |
| 954.550 | 17.6 | | | | 1.81(70)E-29 | | |
| **I-133** | | **20.8(1) h** | | **20.8(1) h** | | **20.8(1) h** | |
| 529.872 | 87 | 1.06(15) d | **2.02(14)E-29** | 21.2(2) h | **6.49(20)E-29** | 1.24(15) d | **1.27(9)E-28** |
| **I-134** | | **52.5(2) m** | | **52.5(2) m** | **1.69(11)E-28** | **52.5(2) m** | **2.62(27)E-28** |
| 847.025 | 95.4 | 1.15(7) h | **5.75(76)E-29** | 1.06(2) h | 1.53(17)E-28 | 1.33(19) h | 2.51(57)E-28 |
| 884.090 | 64.9 | | | 1.11(7) h | 1.59(18)E-28 | | 2.57(32)E-28 |
| 1072.550 | 14.9 | | | 1.12(2) h | 1.94(21)E-28 | | 3.4(13)E-28 |
| 1136.160 | 9.1 | | | | 1.98(33)E-28 | | |
| **I-135** | | **6.57(2) h** | **3.09(14)E-29** | **6.57(2) h** | **7.04(26)E-29** | **6.57(2) h** | **1.72(28)E-28** |
| 1131.511 | 22.7 | 6.8(8) h | 3.11(21)E-29 | 5.7(6) h | 6.99(52)E-29 | 12.9(1) h | 1.77(26)E-28 |
| 1260.409 | 28.9 | 7(5) h | 3.08(18)E-29 | 6.7(4) h | 6.89(28)E-29 | 7(2) h | 1.67(73)E-28 |
| 1791.196 | 7.8 | | | 8.1(16) h | 7.29(70)E-29 | | |
| **Xe-135** | | **9.14(2) h** | | **9.14(2) h** | | **9.14(2) h** | |
| 249.770 | 90 | 16.8(21) h | **2.85(55)E-29** | 16.8(19) h | **7.8(12)E-29** | 20(3) h | **1.95(34)E-28** |
| **Cs-138** | | | | **33.4(2) m** | | **33.4(2) m** | |
| 1435.795 | 76.3 | | | | **1.31(13)E-28** | | **2.03(43)E-28** |
| **La-142** | | **91.1(5) m** | | **91.1(5) m** | **7.73(82)E-29** | **91.1(5) m** | |
| 641.285 | 47 | 1.86(19) h | **3.91(31)E-29** | 1.9(5) h | 8.4(13)E-29 | 2.08(1) h | **1.66(24)E-28** |
| 894.900 | 8.3 | | | 3.46(2) h | 7.3(14)E-29 | | |
| 2397.800 | 13.3 | | | | 7.3(15)E-29 | | |
| **Ce-143** | | **1.38(2) d** | | **1.38(2) d** | | **1.38(2) d** | |
| 293.266 | 42.8 | 1.68(3) d | **3.32(22)E-29** | 1.50(18) d | **7.80(35)E-29** | 3.1(7) d | **2.51(35)E-28** |
| **At-208** | | | | **1.63(3) h** | | **1.63(3) h** | |
| 177.595 | 48.6 | | | | | | **6.1(23)E-28** |
| 845.044 | 19.7 | | | | **3.83(64)E-29** | | |
| **Ac-224** | | | | **2.78(17) h** | | **2.78(17) h** | |
| 131.613 | 26.9 | | | | | | **1.13(51)E-26** |
| 215.983 | 52.3 | | | | **2.84(80)E-29** | | |
| **Pa-233** | | **26.97(1) d** | | **26.97(1) d** | | **26.97(1) d** | |
| 312.170 | 38.6 | 17.8(16) d | **9.55(55)E-27** | | **1.78(3)E-26** | | **4.30(20)E-26** |

## RESULTS OF $^{129}$I

During irradiation the samples $^{129}$I (coated aluminum weighing 17.6 g) and $^{127}$I (in a shell made of plexiglass weighing 2.53 g) were installed on the side of the section №3 uranium assembly Fig.2. In samples $^{129}$I impurity present stable isotope $^{127}$I. To correct account the contribution of $^{127}$I, samples $^{129}$I were irradiated simultaneously with the samples, containing only isotope $^{127}$I.

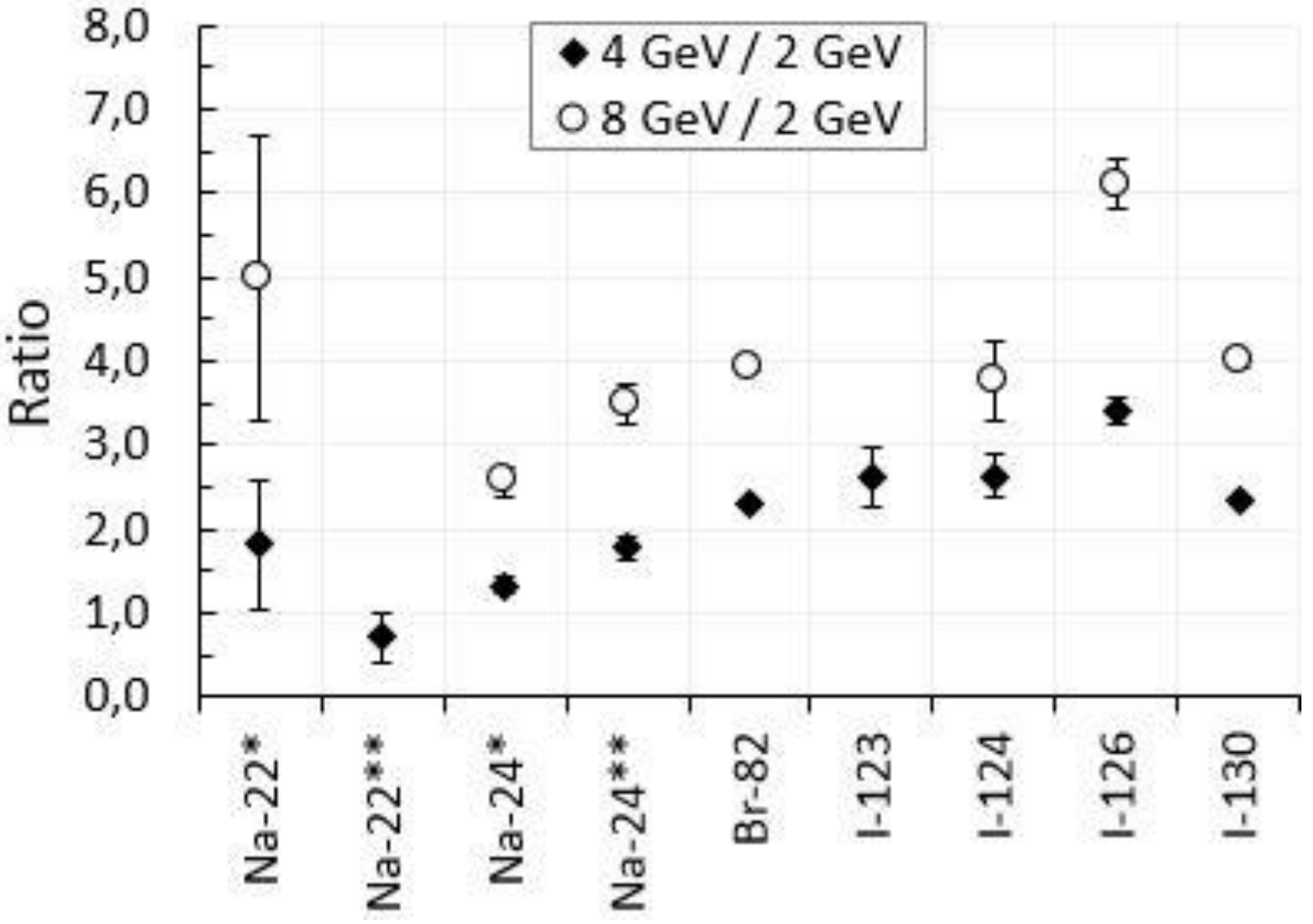

Fig.6. Experimental values of the ratio of reaction rate R(4 GeV) / R(2 GeV) and R(8 GeV) / R(2 GeV) for $^{23}$Na+$^{27}$Al+$^{129}$I with secondary neutrons for product nuclei at energies of deuterons 2, 4, 8 GeV. Na-22* - product $^{23}$Na(n,2n)$^{22}$Na. Na-22** - product $^{27}$Al(n,α2n)$^{22}$Na. Na-24* - product $^{23}$Na(n,γ)$^{24}$Na. Na-24** - product $^{27}$Al(n,α)$^{24}$Na. Values of reaction rates for $^{82}$Br obtained from calculation Calc.2 due to lack of weight values $^{81}$Br.

$^{22}$Na produced simultaneously in two reactions $^{27}$Al(n,α2n)$^{22}$Na ($E_{thr}$=23.35 MэB) and $^{23}$Na(n,2n)$^{22}$Na ($E_{thr}$=12.96 MэB). $^{24}$Na is generated from $^{27}$Al(n,α)$^{24}$Na ($E_{thr}$=3.25 MэB) and $^{23}$Na(n,γ)$^{24}$Na reactions. Share $^{24}$Na generated from the (n,γ) reaction at deuteron energies 2 GeV - 2.1 %, 4 GeV - 0.9 % and 8 GeV - 0.6 %.

$^{22}$Na and $^{24}$Na produced mainly from $^{27}$Al due to large mass of $^{27}$Al. Contribution $^{24}$Na produced from $^{23}$Na calculated using the values of reaction rates for $^{24}$Na in the sample $^{127}$I.

We assume that in the composition of samples $^{129}$I, present admixture of $^{81}$Br and $^{82}$Br is a product of $^{81}$Br(n,γ)$^{82}$Br reaction. The admixture of $^{81}$Br may be, by our estimates (Calc.2), in $^{129}$I no more than 1.5(5)%. $^{82}$Br in the samples $^{127}$I was not observed. $^{123}$I, $^{124}$I, and $^{126}$I are products of (n,7n), (n,6n) and (n,4n) reactions. $^{130}$I is product (n,γ) reaction.

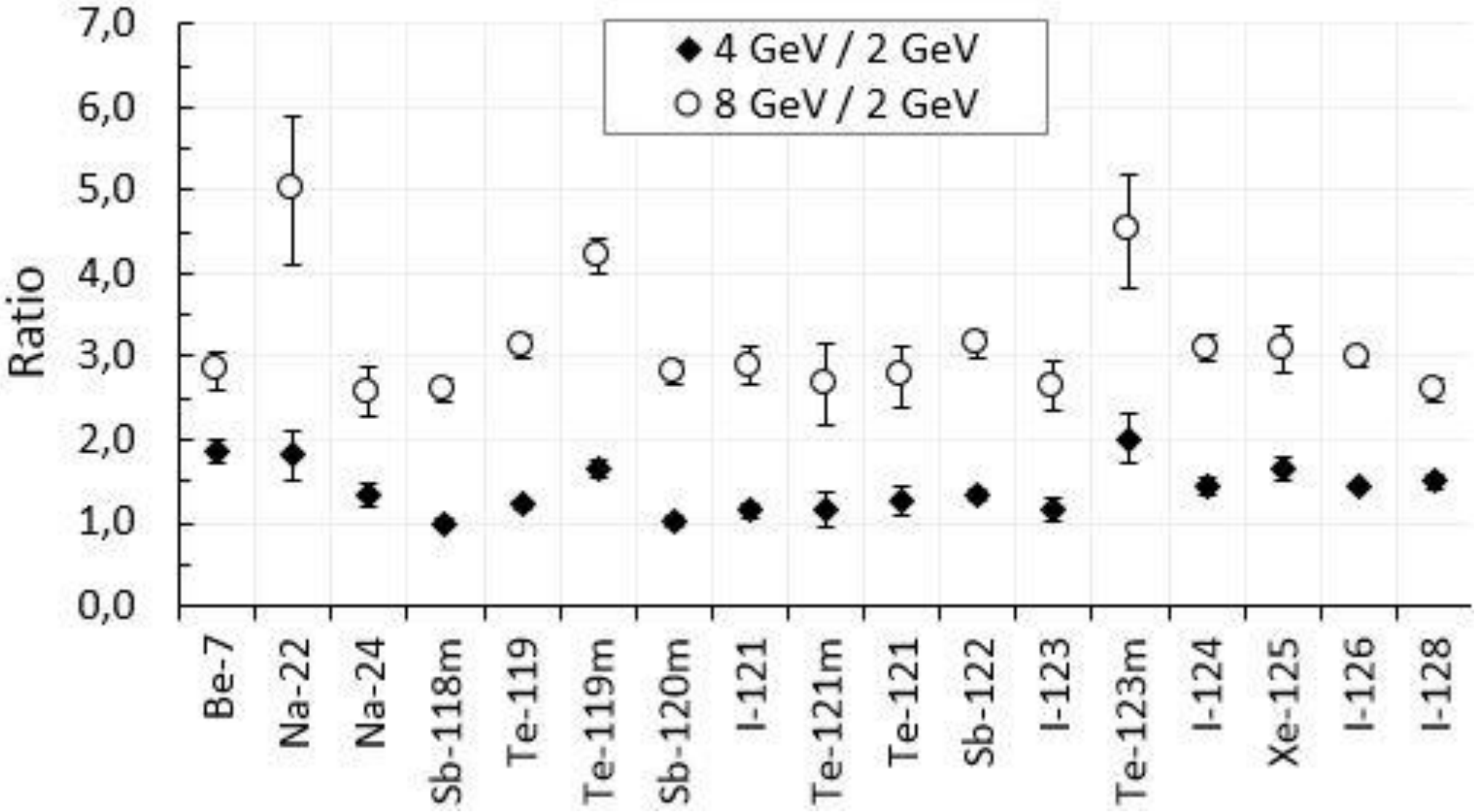

Fig.7. Experimental values of the ratio of reaction rate R(4 GeV) / R(2 GeV) and R(8 GeV) / R(2 GeV) for $^{23}Na$+ $^{127}I$ with secondary neutrons for product nuclei at energies of deuterons 2, 4, 8 GeV.

$^{7}Be$ is produced in the shell of the sample (plexiglass) from $^{12}C$ and $^{16}O$. $^{22}Na$ and $^{24}Na$ produced from $^{23}Na$ in the reactions (n,2n) and (n,γ). $^{118m}Sb$, $^{120m}Sb$ and $^{122}Sb$ are products of (n,α6n) ($E_{thr}$=44.11 MeV), (n,α4n) ($E_{thr}$=27.42 MeV) and (n,α2n) ($E_{thr}$=11.23 MeV) reactions. Radionuclides $^{119}Te$, $^{121}Te$ and $^{123m}Te$ are products of (n,t6n) ($E_{thr}$=57.56 MeV), (n,t4n) ($E_{thr}$=39.92 MeV) and (n,t2n) ($E_{thr}$=23.01 MeV). $^{120}I$, $^{121}I$, $^{123}I$, $^{124}I$, and $^{126}I$ are products of (n,8n) ($E_{thr}$=62.18 MeV), (n,7n) ($E_{thr}$=51.53 MeV), (n,5n) ($E_{thr}$=33.59 MeV), (n,4n) ($E_{thr}$=26.01 MeV) and (n,2n) ($E_{thr}$=44.11 MeV) reactions. $^{128}I$ is product (n,γ) reaction. Table 2 summarizes the results of comparing the experimental and calculated data (Calc.1 and Calc.2) on $^{127}I$ and $^{129}I$.

Table.2. Comparison of results for $^{127}I$ and $^{129}I$ with calculations.

| **Nuclear reactions on $^{127}I$ samples** | **2 GeV** | | **4 GeV** | | **8 GeV** | |
|---|---|---|---|---|---|---|
| | **Exp/Calc. 1** | **Exp/Calc. 2** | **Exp/Calc. 1** | **Exp/Calc. 2** | **Exp/Calc. 1** | **Exp/Calc. 2** |
| **$^{127}I(n,\gamma)^{128}I$** | | 0.64(4) | | 0.52(3) | | 0.53(3) |
| **$^{127}I(n,2n)^{126}I$** | 1.10(4) | 0.66(2) | 0.85(3) | 0.57(2) | 0.69(3) | 0.53(1) |
| **$^{127}I(n,4n)^{124}I$** | 1.18(9) | 0.68(2) | 0.89(8) | 0.64(7) | 0.79(7) | 0.62(4) |
| **Nuclear reactions on $^{129}I$ samples** | | | | | | |
| **$^{129}I(n,\gamma)^{130}I$** | | 0.61(1) | | 0.71(1) | | 0.71(1) |
| **$^{129}I(n,4n)^{126}I$** | 1.45(12) | 0.72(3) | 2.46(17) | 1.56(6) | 1.74(27) | 1.27(6) |
| **$^{129}$I(n,6n)$^{124}$I** | 1.95(26) | 0.88(5) | 2.49(32) | 1.49(19) | 1.43(23) | 0.94(18) |

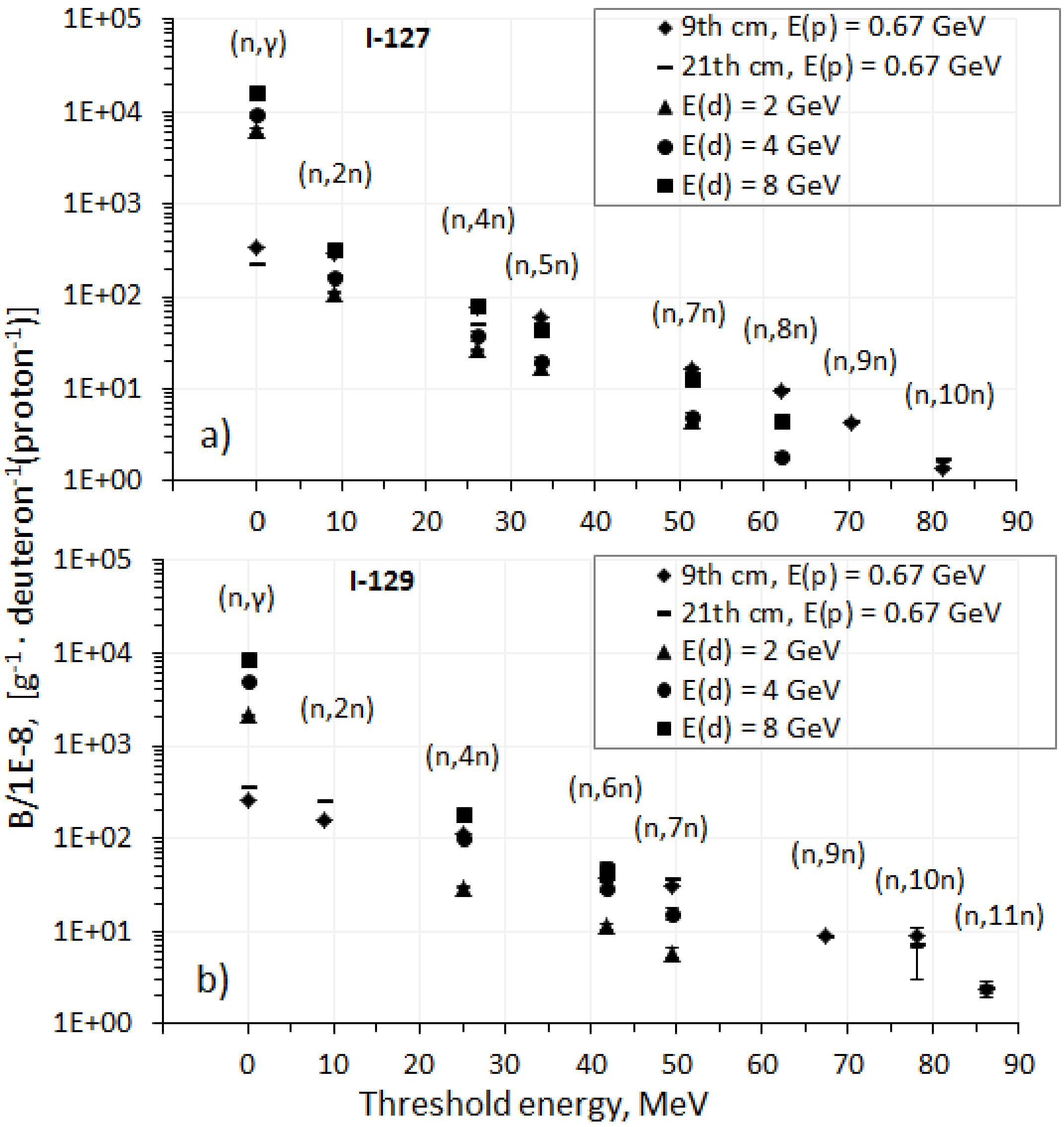

Fig.8. Comparison of the experimental results on $^{127}$I - a) and $^{129}$I - b) with [18].

Previously, we performed an experiment [18], in which for production of secondary neutrons used massive lead assembly installed in the proton beam (660 MeV) in Phasotron DLNP. Data for $^{127}$I and $^{129}$I, obtained in this experiment, we compare with data from experiments at the Nuclotron VBLHEP (see Fig.8). It can be seen that the results for $^{127}$I and $^{129}$I in good agreement for almost all the observed (n,xn) reactions (in this work have not been identified (n, 9n), (n, 10n) reactions in 127I, (n, 9n), (n, 10n) and (n, 11n) reactions in 129I, as the time from end of irradiation up to start of measurement 80 -160 min.), except the (n,γ) reaction, which is caused by the presence on setup "QUINTA" lead shield that reflects neutrons.

Table.3. Values of the reaction rates $^{127}$I with secondary neutrons for product nuclei at energies of deuterons 2, 4, 8 GeV.
(*) denotes mixing due to other nuclide.

| **Isotope**<br>Energy<br>[keV] | $I_g$<br>[%] | **2 GeV** | | **4 GeV** | | **8 GeV** | |
|---|---|---|---|---|---|---|---|
| | | **$T_{1/2}$(Library)**<br>$T_{1/2}$(Exper.) | **<R>**<br>R | **$T_{1/2}$(Library)**<br>$T_{1/2}$(Exper.) | **<R>**<br>R | **$T_{1/2}$(Library)**<br>$T_{1/2}$(Exper.) | **<R>**<br>R |
| **Be-7** | | **53.12(7) d** | | **53.12(7) d** | | **53.12(7) d** | |
| 477.595 | 10.5 | 38(15) d | **3.68(30)E-29** | 25(12) d | **6.81(47)E-29** | | **1.04(9)E-28** |
| **Na-22** | | **2.60(1) y** | | **2.60(1) y** | | **2.60(1) y** | |
| 1274.530 | 99.94 | 100(110) d | **5.58(51)E-29** | | **1.01(24)E-28** | 7.0(15) d | **2.79(75)E-28** |
| **Na-24** | | **14.96(1) h** | **7.22(66)E-29** | **14.96(1) h** | **9.6(11)E-29** | **14.96(1) h** | **1.85(26)E-28** |
| 1368.633 | 100 | 15.18(18) h | 7.04(15)E-29 | 15.21(10) h | 9.40(24)E-29 | 15.17(14) h | 1.79(5)E-28 |
| 2754.028 | 99.94 | 15.31(21) h | 9.65(58)E-29 | 14.86(10) h | 1.39(9)E-28 | 14.76(19) h | 2.93(19)E-28 |
| **Ag-108m** | | | | | | **418(21) y** | **2.70(63)E-30** |
| 433.937 | 90 | | | | | | 2.57(73)E-30 |
| 614.276 | 89.8 | | | | | | 2.85(76)E-30 |
| 722.907 | 90.8 | | | | | | 6.9(13)E-30* |
| **In-111** | | **2.80(1) d** | **5.87(72)E-31** | | | | |
| 171.280 | 90 | 2.8(4) d | 5.26(40)E-31 | | | | |
| 245.395 | 94 | 2.5(4) d | 6.73(47)E-31 | | | | |
| **Sb-118m** | | **5.00(2) h** | **1.55(6)E-30** | **5.00(2) h** | **1.54(11)E-30** | **5.00(2) h** | **4.01(30)E-30** |
| 253.678 | 99 | 5.09(25) h | 1.47(7)E-30 | 5.1(4) h | 1.44(9)E-30 | 5.0(4) h | 3.44(21)E-30 |
| 1050.650 | 97 | 5.8(7) h | 1.64(11)E-30 | 4.6(7) h | 1.76(16)E-30 | 4.7(4) h | 4.51(35)E-30 |
| 1229.680 | 100 | 5.1(4) h | 1.62(10)E-30 | 8.3(14) h | 1.73(23)E-30 | 6.2(8) h | 4.26(41)E-30 |
| **Te-119** | | **16.03(5) h** | | **16.03(5) h** | | **16.03(5) h** | |
| 644.010 | 84 | 19.0(6) h | **3.24(14)E-30** | 16.9(7) h | **3.92(13)E-30** | 15.5(6) h | **1.10(4)E-29** |
| **Te-119m** | | **4.70(4) d** | **2.49(16)E-30** | **4.70(4) d** | **4.09(23)E-30** | **4.70(4) d** | **1.05(4)E-29** |
| 153.590 | 66 | 4.9(4) d | 2.39(11)E-30 | 4.1(6) d | 4.20(36)E-30 | 3.92(23) d | 1.02(5)E-29 |
| 1212.730 | 66 | 5.4(7) d | 2.75(17)E-30 | 11.3(28) d | 4.02(30)E-30 | 4.6(8) d | 1.11(8)E-29 |
| **I-120** | | | | **81.0(6) m** | | **81.0(6) m** | |
| 560.440 | 73 | | | 1.32(2) h | **3.87(36)E-30** | 1.40(14) h | **9.3(8)E-30** |
| **Sb-120m** | | **5.76(2) d** | **1.58(8)E-30** | **5.76(2) d** | **1.61(19)E-30** | **5.76(2) d** | **4.43(24)E-30** |
| 197.300 | 87 | 12.3(21) d | 1.50(13)E-30 | 5.5(7) d | 1.51(21)E-30 | 5.6(5) d | 3.83(33)E-30 |
| 1023.100 | 99.4 | 9(4) d | 1.63(16)E-30 | 1.6(10) d | 1.86(57)E-30 | 5.4(5) d | 4.44(33)E-30 |
| 1171.300 | 100 | 5.4(13) d | 1.63(12)E-30 | | | 5.0(15) d | 4.52(92)E-30 |
| **I-121** | | **2.12(1) h** | | **2.12(1) h** | | **2.12(1) h** | |
| 212.189 | 84 | 2.14(9) h | **9.48(73)E-30** | 2.05(10) h | **1.07(7)E-29** | 1.99(9) h | **2.71(21)E-29** |
| **Te-121m** | | **154(7) d** | | **154(7) d** | | **154(7) d** | |
| 212.189 | 81 | | **8.7(16)E-30** | | **1.01(20)E-29** | | **2.33(42)E-29** |
| **Te-121** | | **16.78(35) d** | **1.26(26)E-29** | **16.78(35) d** | | **16.78(35) d** | **3.47(20)E-29** |
| 507.591 | 17.7 | 23(8) d | 2.37(22)E-29 | | | 13(5) d | 6.23(73)E-29* |
| 573.139 | 80.3 | 22.7(10) d | 1.20(5)E-29 | 19.3(13) d | **1.60(11)E-29** | 16.6(19) d | 3.47(20)E-29 |
| **Sb-122** | | **2.72(1) d** | | **2.72(1) d** | | **2.72(1) d** | |
| 564.119 | 71 | 2.64(23) d | **2.29(11)E-30** | 2.4(3) d | **3.08(16)E-30** | 2.89(22) d | **7.19(36)E-30** |
| **I-123** | | **13.27(8) h** | | **13.27(8) h** | | **13.27(8) h** | |
| 158.970 | 83 | 13.0(6) h | **3.59(41)E-29** | 12.3(6) h | **4.18(45)E-29** | 12.5(7) h | **9.5(11)E-29** |
| **Te-123m** | | **119.7(1) d** | | **119.7(1) d** | | **119.7(1) d** | |
| 158.970 | 84 | | **7.7(13)E-30** | | **1.55(21)E-29** | | **3.47(47)E-29** |
| **I-124** | | **4.18(2) d** | **5.54(18)E-29** | **4.18(2) d** | **7.93(85)E-29** | **4.18(2) d** | **1.71(12)E-28** |
| 602.729 | 63 | 4.18(15) d | 5.56(15)E-29 | 4.13(5) d | 7.29(30)E-29 | 3.96(9) d | 1.56(5)E-28 |
| 722.786 | 10.3 | 4.14(22) d | 5.26(17)E-29 | 17(7) d | 1.21(10)E-28 | 5.1(11) d | 2.44(21)E-28 |
| 1325.512 | 1.6 | 3.4(10) d | 7.40(81)E-29 | 4.8(28) d | 1.03(10)E-28 | 3.4(3) d | 2.46(20)E-28 |
| 1509.470 | 3.1 | | | 5.5(7) d | 6.9(12)E-29 | 4.5(5) d | 1.97(22)E-28 |
| 1690.983 | 10.9 | 4.6(4) d | 5.84(23)E-29 | 4.9(3) d | 7.02(39)E-29 | 4.07(17) d | 1.63(6)E-28 |
| **Xe-125** | | **16.9(2) h** | **7.93(70)E-31** | **16.9(2) h** | **1.31(10)E-30** | **16.9(2) h** | **2.45(22)E-30** |
| 188.418 | 54 | 17.1(26) h | 7.57(61)E-31 | 15.3(23) h | 1.28(10)E-30 | 20(6) h | 2.10(20)E-30 |
| 243.378 | 30 | 1.8(5) d | 9.3(12)E-31 | | 1.61(30)E-30 | | 2.89(63)E-30 |
| **I-126** | | **13.11(5) d** | **2.32(7)E-28** | **13.11(5) d** | **3.35(12)E-28** | **13.11(5) d** | **6.90(18)E-28** |
| 388.633 | 34.1 | 13.3(6) d | 2.49(5)E-28 | 12.37(16) d | 3.44(16)E-28 | 10.7(5) d | 6.85(22)E-28 |
| 491.243 | 2.8 | 12.8(6) d | 2.33(8)E-28 | 11.8(6) d | 3.27(18)E-28 | 11.1(11) d | 6.44(28)E-28 |
| 666.331 | 33.1 | 13.8(7) d | 2.27(5)E-28 | 11.98(29) d | 3.36(11)E-28 | 10.8(6) d | 6.73(16)E-28 |
| 753.819 | 4.2 | 14.0(7) d | 2.16(6)E-28 | 11.5(6) d | 3.14(17)E-28 | 11.0(11) d | 6.53(25)E-28 |
| 879.876 | 0.7 | 15(4) d | 2.57(29)E-28 | 12.8(18) d | 4.74(48)E-28 | 9.1(12) d | 9.68(97)E-28* |
| **I-128** | | **24.99(2) m** | | **24.99(2) m** | **1.95(12)E-26** | **24.99(2) m** | **3.38(17)E-26** |
| 442.901 | 17 | 25.2(1) m | **1.30(8)E-26** | 25.06(2) m | 1.97(12)E-26 | 24.90(2) m | 3.45(6)E-26 |
| 526.557 | 1.6 | | | 31.1(1) m | 1.87(26)E-26 | 24.68(2) m | 3.29(8)E-26 |

Table.4. Values of the reaction rates [129]I with secondary neutrons for product nuclei at energies of deuterons 2, 4, 8 GeV.
(*) denotes mixing due to other nuclide.

| **Isotope**<br>Energy<br>[keV] | $I_g$<br>[%] | **2 GeV**<br>**$T_{1/2}$(Library)**<br>$T_{1/2}$(Exper.) | <br>**<R>**<br>R | **4 GeV**<br>**$T_{1/2}$(Library)**<br>$T_{1/2}$(Exper.) | <br>**<R>**<br>R | **8 GeV**<br>**$T_{1/2}$(Library)**<br>$T_{1/2}$(Exper.) | <br>**<R>**<br>R |
|---|---|---|---|---|---|---|---|
| **Na-22*** | | **2.60(1) y** | | **2.60(1) y** | | **2.60(1) y** | |
| 1274.530 | 99.94 | 40(24) d | **5.6(34)E-29** | 40(24) d | **1.01(23)E-28** | | **2.79(20)E-28** |
| **Na-22**** | | **2.60(1) y** | | **2.60(1) y** | | **2.60(1) y** | |
| 1274.530 | 99.94 | 40(24) d | **8.1(49)E-30** | 40(24) d | **5.7(13)E-30** | | **6.42(46)E-28** |
| **Na-24*** | | **14.96(1) h** | **7.22(43)E-29** | **14.96(1) h** | **9.60(79)E-29** | **14.96(1) h** | **1.85(14)E-28** |
| 1368.633 | 100 | 15.02(3) h | | 15.6(4) h | | 15.0(4) h | |
| 2754.028 | 99.94 | 15.05(3) h | | 15.6(5) h | | 14.9(5) h | |
| **Na-24**** | | **14.96(1) h** | **2.72(16)E-29** | **14.96(1) h** | **4.83(40)E-29** | **14.96(1) h** | **9.47(74)E-29** |
| 1368.633 | 100 | 15.02(3) h | | 15.6(4) h | | 15.0(4) h | |
| 2754.028 | 99.94 | 15.05(3) h | | 15.6(5) h | | 14.9(5) h | |
| **Br-82** | | **35.30(2) h** | **5.09(13)E-29** | **35.30(2) h** | **6.37(12)E-28** | **35.30(2) h** | **1.11(4)E-27** |
| 554.348 | 70.8 | 1.08(5) d* | 6.90(77)E-29 | 1.45(5) d | 6.82(20)E-28 | 1.25(5) d | 1.17(10)E-27 |
| 619.106 | 43.4 | 1.42(3) d | 4.70(12)E-29 | 1.58(6) d | 5.94(22)E-28 | 1.33(5) d | 1.02(6)E-27 |
| 698.374 | 28.5 | 1.44(4) d | 4.64(14)E-29 | 1.68(7) d | 6.06(30)E-28 | 1.74(20) d | 1.35(8)E-27 |
| 776.517 | 83.5 | 1.45(2) d | 5.02(9)E-29 | 1.56(6) d | 6.02(21)E-28 | 1.39(4) d | 1.09(5)E-27 |
| 827.828 | 24 | 1.45(5) d | 5.43(17)E-29 | 1.53(7) d | 6.30(24)E-28 | 1.41(2) d | 0.99(10)E-27 |
| 1044.002 | 27.2 | 1.48(5) d | 5.31(18)E-29 | 1.52(6) d | 6.55(25)E-28 | 1.7(3) d | 1.13(11)E-27 |
| 1317.473 | 26.5 | 1.54(5) d | 5.61(17)E-29 | 1.61(7) d | 6.59(28)E-28 | 1.32(11) d | 1.15(9)E-27 |
| 1474.880 | 16.3 | 1.45(6) d | 5.39(20)E-29 | 1.58(6) d | 6.56(26)E-28 | 1.36(12) d | 1.11(8)E-27 |
| **I-123** | | **13.27(8) h** | | **13.27(8) h** | | | |
| 158.970 | 83 | 12.3(6) h | **1.26(16)E-29** | 10(3) h | **3.30(51)E-29** | | |
| **I-124** | | **4.18(2) d** | **2.42(15)E-29** | **4.18(2) d** | | **4.18(2) d** | |
| 602.729 | 63 | 3.9(7) d | 2.32(16)E-29 | 5.9(13) d | **6.37(81)E-29** | 5.0(17) d | **9.1(17)E-29** |
| 1690.983 | 10.9 | 4.1(19) d | 2.81(33)E-29 | | | | |
| **I-126** | | **13.11(5) d** | **6.42(28)E-29** | **13.11(5) d** | **2.19(9)E-28** | **13.11(5) d** | **3.92(20)E-28** |
| 388.633 | 34.1 | 3.8(12) y | 6.66(27)E-29 | 27(12) d | 2.23(10)E-28 | | 4.13(25)E-28 |
| 666.331 | 33.1 | 4.4(19) y | 6.05(30)E-29 | 26(12) d | 2.14(10)E-28 | | 3.68(28)E-28 |
| 753.819 | 4.2 | 17(7) d | 4.54(56)E-29 | | | | |
| **I-130** | | **12.36(3) h** | **4.59(10)E-27** | **12.36(3) h** | **1.07(2)E-26** | **12.36(3) h** | **1.84(4)E-26** |
| 418.010 | 34.2 | 12.47(2) h | 4.15(7)E-27 | 12.8(4) h | 1.01(4)E-26 | 12.27(29) h | 1.76(6)E-26 |
| 457.720 | 0.2 | 10.1(18) h | 3.98(39)E-27 | | | | |
| 536.090 | 99 | 12.44(2) h | 4.48(7)E-27 | 12.8(4) h | 1.05(4)E-26 | 12.2(4) h | 1.81(6)E-26 |
| 539.100 | 1.4 | 12.49(17) h | 4.48(11)E-27 | 12.43(27) h | 1.12(3)E-26 | 13.2(9) h | 1.93(13)E-26 |
| 553.900 | 0.7 | 1.08(5) d* | 8.31(8)E-27* | | | | |
| 586.050 | 1.7 | 12.51(12) h | 4.34(10)E-27 | 12.9(5) h | 1.03(4)E-26 | 12.5(7) h | 1.70(8)E-26 |
| 603.500 | 0.6 | 13.8(5) h | 5.35(20)E-27 | 13.4(18) h | 1.13(9)E-26 | 2.37(15) d* | 2.12(22)E-26 |
| 668.540 | 96 | 12.45(2) h | 4.48(7)E-27 | 12.9(4) h | 1.05(4)E-26 | 12.3(5) h | 1.79(6)E-26 |
| 685.990 | 1.1 | 12.53(16) h | 4.33(10)E-27 | 13.1(5) h | 0.98(4)E-26 | 12.9(18) h | 1.91(13)E-26 |
| 739.480 | 82 | 12.42(2) h | 4.43(7)E-27 | 12.9(4) h | 1.06(4)E-26 | 12.1(4) h | 1.80(6)E-26 |
| 800.230 | 0.1 | 11.6(19) h | 4.17(43)E-27 | | | | |
| 808.290 | 0.2 | 13.0(6) h | 4.22(20)E-27 | 14.9(20) h | 0.92(8)E-26 | | |
| 877.350 | 0.2 | 11.5(10) h | 4.93(36)E-27 | | | | |
| 967.020 | 0.9 | 12.47(24) h | 4.51(11)E-27 | 13.3(8) h | 1.08(5)E-26 | | 2.21(24)E-26 |
| 1096.480 | 0.5 | 1.5(5) d | 4.80(9)E-27 | 13.7(11) h | 1.21(7)E-26 | | 2.34(35)E-26 |
| 1122.150 | 0.2 | 12.7(10) h | 5.60(30)E-27 | 16.4(25) h | 1.44(19)E-26 | | |
| 1157.470 | 11.3 | 12.37(4) h | 5.31(10)E-27 | 13.1(5) h | 1.11(4)E-26 | 11.8(4) h | 1.97(6)E-26 |
| 1222.560 | 0.2 | 12.5(8) h | 5.54(27)E-27 | | | | |
| 1272.120 | 0.7 | 12.65(24) h | 5.67(13)E-27 | 13.6(7) h | 1.24(7)E-26 | 10(4) h | 2.45(30)E-26 |
| 1403.900 | 0.3 | 12.3(3) h | 5.08(16)E-27 | 14.1(20) h | 1.08(12)E-26 | | |

## CONCLUSIONS

Is interesting to compare the results for $^{232}$Th of this work with the results of other experiments. In [33] on setup «GAMMA-2»: Pb target (d = 8 cm, l = 20 cm), paraffin moderator (6 cm thick). $^{232}$Th sample located on the end surface of the target. Installation was irradiated by protons with an energy of 1 GeV. The reaction rate (n,γ) R = 9.80(6)E-27, almost equal to the results of this work at a deuteron energy 2 GeV. In [9] on the setup "Energy plus Transmutation": overall length 48 cm (4 sections), the length of Pb target in four sections of 45.6 cm and a diameter of 8.4 cm target surrounded by uranium blanket thickness of two elements (see present work) and irradiated with deuterons 1.6 GeV. Samples located on the surface of the blanket. The reaction rate (n,γ) R = 3.03(10)E-26 and fission (n,f) R = 5.89(60)E-27 - 3 and 10 times higher than results of this work at 2 GeV. Also interesting comparison with the work [17], made at a deuteron energy of 2.33 GeV at the setup "GAMMA-3": a lead target (d = 8 cm, l = 60 cm) in the carbon moderator with large size, channels for samples at different distances from the center of the target. Results for R - (n,γ) in a yield of 233Pa (will eventually 233U) is 30-40 times higher and fission (n, f) - 20 - 30 times higher than present work for the nearest window on carbon (24 cm from the axis target). This is due to the high ability of carbon to slowing of spallation neutrons coming from the target.

Ratio of the weight of produced $^{233}$U to $^{232}$Th at 2 GeV - 2.90(17)E-13 at 4 GeV 4.88(8)E-13 and at 8 GeV 3.93(18)E-13. Ratio of the weight of produced $^{130}$I to $^{129}$I in the conditions of the present experiment at 2 GeV 1.40(3)E-13 at 4 GeV 2.94(6)E-13 and at 8 GeV 1.69(4)E-13. If we calculate of transmutation for conditions: 10 mA and irradiate for 30 days. It is: 2 GeV – 0.082(2)%; 4 GeV – 0.190(4)% and 8 GeV – 0.330(7)%. These calculations allow us to estimate the transmutation at high currents and irradiation time - in the tens of percent. By Calc.1 and experimental data 90(5)% of the transmutation $^{129}$I is (n,γ) reaction at all three deuteron energies (2, 4, 8 GeV).

These results are interesting to compare with those obtained previously. The first work was done with $^{129}$I in the direct beam of protons Phasotron JINR $E_p$ = 660 MeV and a current of 1.2 mA [34]. Samples with $^{129}$I and $^{127}$I were in direct proton beam. The paper presents number: 30 days and 10 mA. However, during its obtaining were taken coefficient "2" - taking into account the formation of stable and long-lived isotopes, as well as registration of gamma rays with energies lower than 300 keV (samples $^{129}$I are measured with a filter: Pb – 10 mm, Cd – 2 mm and Cu – 1 mm). Therefore, comparisons should take a number (2-3)%. It should be borne in mind that the experimental conditions: the direct beam of protons and not heavy target surrounded uranium, carbon or paraffin moderator - significantly different from the next. In experiments with beams of relativistic protons nuclotron JINR VBLHEP with energies: 1.0, 1.5 and 2.0 GeV [10] - transmutation is respectively, - 0.075, 0.132 and 0.153% at a current of 10 mA and 30 days of irradiation, which is close to the results of this work. In a similar study with deuterons 2.52 GeV [8], a transmutation of 0.132%. In [11] the results for protons with energy 2.0 GeV - transmutation is 0.13%. In [35, 36] at the proton beam with the energy of 3.67 GeV (Nuclotron JINR), lead target (d = 8 cm, l = 20cm, paraffin moderator) transmutation of $^{129}$I was 0.9% (10 mA, 30 days). Grows of transmutation is associated from energy, obviously with increase the yield of spallation neutrons from the target (70 ± 20%) and including - thermal, making the main contribution to the reaction (n,$\gamma$) - more than 92%.